\documentclass[12pt, letterpaper]{article}

\usepackage{amsmath, amsfonts, amscd, amssymb}
\usepackage{amsthm}
\usepackage[all,cmtip]{xy}

\newcommand{\aconn}{\mathcal{A}}

\newcommand{\aut}{{\mathcal{A}}ut}

\newcommand{\conn}{\mathcal{D}}

\newcommand{\cons}{\mathbf{C}}

\newcommand{\curv}{R}

\newcommand{\modl}{\mathbf{\mathcal{E}}}

\newcommand{\smooth}{\mathcal{C}^{\infty}}

\newcommand{\struc}{\mathbf{A}}

\newcommand{\triad}{{\mathfrak{T}}}

\newcommand{\R}{\mathbb{R}}

\title{\bf A Heuristic Philosophical Discourse on Various Applications of \\ Abstract Differential Geometry in Quantum Gravity Research\thanks{This paper is a philosophical distillation of the basic concepts and central results of this author's ongoing research project, spanning the last three decades, of applying Anastasios Mallios's Abstract Differential Geometry to the `persistently stubborn' problem of formulating a conceptually sound, mathematically consistent and calculationally finite Quantum Theory of Gravity. The paper is posted at the {\bf General Relativity and Quantum Cosmology} website {\it www.arXiv.org/gr-qc}, as follows: {\bf gr-qc/2408.07725}. This work may be viewed as the philosophical {\it r\'esum\'e} and aftermath (:after the Maths!) of the following published papers \cite{malrap1, malrap2, malrap3, rap1, rap2, rap5, rap7, rap11, rap13, rap14, rapzap1, rapzap2}, as well as a sequel to the {\em Dodecalogue} paper \cite{rap14} in the aftermath of the recent publication \cite{rap15} and the pre-prints \cite{rap16, rap17, rap18}, which are currently work-in-progress in the pipeline. In turn, a longer version of the paper will constitute a chapter in a research monograph type of book that we had and have been working on, in collaboration with the late Professor Anastasios Mallios, since 2003 \cite{malrap4}.}}

\author{Ioannis Raptis\thanks{Supply \& Substitute Secondary School Teacher of Mathematics, Physics and Chemistry, Reeson Education, London, United Kingdom; email: {\it irapti11@gmail.com}}}

\date{Tuesday, 20th of August 2024\\ {\bf (Version II)}}

\begin{document}

\maketitle

\pagestyle{myheadings}\markboth{\centerline {\small {\sc
{Ioannis Raptis}}}}{\centerline
{\footnotesize {\sc {Philosophical Aspects of ADG in QG}}}}

\pagenumbering{arabic}

\begin{abstract}

\noindent{\small In the present paper, we outline and expound the fundamental and novel qualitative-cum-philosophical premises, principles, ideas, concepts, constructions and results that originate from our ongoing research project of applying the new conceptual panoply and the novel technical machinery of Abstract Differential Geometry (ADG) to various persistently outstanding issues in Quantum Gravity (QG) \cite{malrap1, malrap2, malrap3, rap1, rap2, rap5, rap7, rap11, rap13, rap14, rapzap1, rapzap2}. This paper may be regarded as a sequel to the paper \cite{rap14} in the aftermath of the paper \cite{rap15}, which is currently in press. At the end of the paper, we discuss the potential philosophical repercussions of two possible future research routes that the main stream of our applications of ADG to QG may bifurcate towards in view of three independent, but overlapping, research papers that are currently under development \cite{rap16, rap17, rap18}.}
\vskip 0.1in



\end{abstract}

\newpage

\section{Prolegomena-cum-Motivation: Why Adopt a Philosophical Stance in Quantum Gravity Research?}

Quantum Gravity (QG), very broadly speaking, is an attempt to unite the laws of Physics that describe dynamics at large (:cosmological) scales---as encoded in Einstein's General Relativity (GR) equations for the gravitational field that guides the motion of large material objects, with the dynamical laws for the other three fundamental forces that guide matter fields and their quantum particles at small (:subatomic) scales---as encoded in Quantum Theory and its application to Special Relativistic field physics, commonly known as Quantum Field Theory (QFT).\footnote{In this paper, we use the terms QFT and {\em Quantum Gauge} (QGT) or {\em Quantum Yang-Mills Theories} (QYMT) of matter interchangeably.}

There is currently a plethora of various and glaringly diverse approaches to QG, and it is not the `vain' or `quixotic' aim of this paper to list them all herein, let alone to review their conceptual and technical import or their successes and shortcomings. On the one hand, there is no unanimous agreement on what the `right' or `correct' approach to QG is and the diversity of the various different approaches---from string theory \cite{gsw,senzwiebach},\footnote{For current developments in this field, see the second reference above \cite{senzwiebach}.} to loop quantum gravity \cite{ash,ash0,ash1,loop}\footnote{For an exhaustive modern exposition of loop quantum gravity, see the last reference \cite{loop} above.} and the `discrete' causal sets \cite{bomb87,sork4,sork1,sork2}, for example---exemplifies exactly that. 

On the other hand, it would be informative for the reader to list a triplet of very general predicates and characteristics---{\em theory making `imperatives'} as it were---that the desired and hitherto elusive Quantum Theory of Gravity should possess, properties that all the aforementioned diverse approaches to QG aim or aspire to satisfy in one way or another:

\begin{enumerate}

\item {\bf Economy (E):} {\em Conceptual clarity and depth, as well as economy and simplicity of the underlying fundamental physical theory construction principles};

\item  {\bf Mathematical Consistency (M):} {\em Internal mathematical consistency (:self-consistency), mathematical representability and predictive power, technical innovation and efficacy, wide ranging utility and versatility of application, and, as a bonus, abstract mathematical simplicity and beauty};\footnote{With respect to this last point, we tacitly assume that`{\em beauty}, especially of the mathematical kind, {\em is in the eye of the beholder}'---in this case, of the theoretician/mathematician---hence it is largely of a {\em subjective aesthetic nature}.}

\item  {\bf Finiteness (F):} {\em Calculational freedom from unphysical (observable) infinities, anomalies and `singularities' of all kinds at which the aforementioned physical laws might seem to break down, hence are deemed to become unphysical, uninformative, unpredictive and, ultimately, practically useless and therefore obsolete}.

\end{enumerate}

\noindent We will henceforth refer to the triplet of desirable QG traits above by the acronym \underline{\bf EMF1}.

\vskip 0.1in

\begin{quotation}

\noindent {\bf `Mission Statement':} Our philosophical discourse in the present paper will focus on showing, arguing and discussing, with extensive references to the existing published literature, that the ADG-theoretic approach to QG, hitherto to be referred to as {\em ADG-gravity}, goes a long way towards satisfying the {\bf EMF1} triptych of theory making imperatives above.

\end{quotation}

\vskip 0.1in

At the same time, the basic `justification' for engaging in a philosophical discourse about the import of ADG-gravity in QG research is another triplet \underline{\bf EMF2}:

\begin{enumerate}

\item {\bf Explication (E):} {\em Explanation and interpretation of new concepts, techniques and results from applying ADG in QG research};

\item  {\bf Mathematical Efficacy (M):} {\em Discussion of the mathematical power of ADG in addressing and resolving certain key QG problems and issues associated with EMF1};

\item {\bf Future Prospects and Developments (F):} {\em Discussion of future prospects for QG theory growth and development, as well as anticipation in what direction will QG research move in view of the new ideas and theoretical paradigms that ADG brings forth.}

\end{enumerate}

\noindent \underline{\bf Expository Declaration 1:} In the sections that follow, whenever we discuss and analyse a qualitative-philosophical aspect of ADG-gravity, at the end of the discussion we will mark it by boldface markers in brackets like, for example, ({\bf F1}): this would mean that the qualitative/philosophical characteristic of ADG-gravity being analysed and discussed  satisfies the {\bf Finiteness 1} aspect of the {\bf EMF1} triplet outlined above. 

\vskip 0.1in

We would like to kick-off our philosophical discourse on applications of ADG in QG with a very telling quote of Gerard 't Hooft just after the turn of the new millennium \cite{thooft}, which kind of gives a {\it raison d'\^etre} and a {\it raison de faire} to our endeavours herein:

\bigskip\noindent (Q1)\hskip 0.9in
\begin{minipage}{11cm}
\noindent ``{\small ...The problems of quantum gravity are much
more than purely technical ones. {\em They touch upon very essential
philosophical issues}\footnote{Our emphasis.}...}''
\end{minipage}

\noindent the basic idea behind the quotation above is that it motivates us to go beyond the `technicalities' of various formal conceptual and mathematical issues in current QG research, and discuss the deeper semantics and philosophical nature underlying, or even possibly transcending, those conceptual and mathematical `technicalities'. Which brings us to the second Expository Declaration of this paper.

\vskip 0.07in

\noindent \underline{\bf Expository Declaration 2:} It is a conscious decision and choice of this author not to include a single quantitative expression ({\it e.g.}, technical mathematical formula) in the present paper, which is purely of a qualitative (discursive) character. For concise definitions of formal technical concepts and their associated mathematical formulae/modelling/equations and associated calculations using ADG, the reader will be directly referred to the relevant published literature.

\vskip 0.07in

\noindent \underline{\bf Expository Declaration 3:} Our philosophical treatise below is organised in short sections that end with `{\em Aphorisms}'---short and terse statements that distill the main philosophical gist of each section.

\section{General Relativity is formulated by the Classical Differential Geometry on a Pointed Smooth Base Spacetime Manifold}

$\bullet$ General Relativity (GR), the classical theory of gravity, is inextricably tied to a $\smooth$-smooth ({\it alias}, differential) base spacetime manifold $M$ for its mathematical formulation via Classical Differential Geometry (CDG)---the Newtonian Calculus based geometry of differential manifolds.

$\bullet$ In turn, the differential manifold $M$, as a geometrical point-set, is equivalent to the {\em structure algebra sheaf} $\smooth(M)$ of germs of smooth coordinate functions of its points.\footnote{This is an example of the general notion of {\em Gelfand Duality} that is one of the fundamental pillars on which ADG was originally erected by Mallios \cite{mall1,mall2,mall4} and used extensively in its applications to QG thereafter \cite{malrap1,malrap2,malrap3,rap5,rap7,rap13,rap14}, and very recently in \cite{rap15}.}

$\bullet$ In GR, the dynamical law of gravity is formulated as Einstein's nonlinear partial differential equations of a $\smooth$-smooth spacetime metric $g_{\mu\nu}$ (and its derivatives), whose $10$ components are supposed to represent the smooth gravitational field potentials.

$\bullet$ Accordingly, GR's Priciple of General Covariance (PGC) is mathematically represented by the group $\mathrm{Diff}(M)$ of (active) {\em diffeomorphisms} of $M$.\footnote{By definition, a diffeomorphism of a smooth manifold $M$ is an automorphism of $M$ that preserves its differential geometric (:smooth) structure, as the latter is effectively encoded in the {\em structure algebra sheaf} $\smooth(M)$ of smooth coordinate functions of $M$'s point events, as noted above.}

\begin{quotation}

\noindent \underline{\bf Aphorism 1.} GR uses CDG to formulate the gravitational dynamics as a differential equation for the smooth metric (and its derivatives) on a background geometrical differential spacetime manifold $M$. In turn, the PGC of GR is represented by the spacetime diffeomorphism group $\mathrm{Diff}(M)$ of the underlying $\smooth$-smooth manifold $M$.

\end{quotation}

\subsection{Feynman's `Fancy-Schmanzy' Differential Geometry and Isham's No-Go of Differential Geometry in Quantum Gravity}

Arguably, the smooth background geometrical spacetime manifold, whether curved or flat, is responsible for both the singularities of the smooth gravitational field $g_{\mu\nu}$ of GR \cite{geroch1,geroch2,hawk1,hawk2,hawk3,hawk4,mtw}, as well as for the pestilential non-renormalisable unphysical infinities that assail QFT on flat Minkowski spacetime \cite{haag,au}.({\bf F1})

Mainly due to these pathologies, QG researchers as early as Feynman and as recently as Isham, have questioned altogether the use of CDG on a smooth spacetime manifold, whether flat (QFT) or curved (GR), as the appropriate mathematical framework via which to formulate QG.

In this line of thought, we first recall from Bryan Hatfield's {\em Quantum Gravity} Foreword to {\em Feynman's Lectures on Gravitation} \cite{feyn} the following telling excerpt:    

\bigskip\noindent (Q2)\hskip 0.9in
\begin{minipage}{11cm}
\noindent ``...Thus it is no surprise that Feynman would
recreate general relativity from a non-geometrical viewpoint. The
practical side of this approach is that one does not have to learn
some `{\em fancy-schmanzy}' (as he liked to call it) differential
geometry in order to study gravitational physics. (Instead, one
would just have to learn some quantum field theory.) However, when
the ultimate goal is to quantize gravity, Feynman felt that the
geometrical interpretation just stood in the way. From the field
theoretic viewpoint, one could avoid actually defining---up
front---the physical meaning of quantum geometry, fluctuating
topology, space-time foam, {\it etc.}, and instead look for the
geometrical meaning after quantization... Feynman certainly felt
that the geometrical interpretation is marvellous, `{\em but the
fact that a massless spin-$2$ field can be interpreted as a metric
was simply a coincidence that might be understood as representing
some kind of gauge invariance'}\footnote{Our emphasis of Feynman's
words as quoted by Bryan Hatfield.}...''
\end{minipage}

\noindent And he further added categorimatically in \cite{feyn3} that:

\bigskip\noindent (Q3)\hskip 0.9in
\begin{minipage}{11cm}
\noindent ``{\em ...the simple ideas of [differential] 
geometry, extended down to infinitely small, are wrong.}''
\end{minipage}

\vskip 0.1in

While more recently, Chris Isham firmly posited in \cite{ish}: 

\bigskip\noindent (Q4)\hskip 0.9in
\begin{minipage}{11cm}
\noindent ``{\em ...at the Planck-length
scale, differential geometry is simply incompatible with quantum
theory...[so that] one will not be able to use differential
geometry in the true quantum-gravity theory...}''
\end{minipage}

\vskip 0.1in

We may distill the above to our second Aphorism:

\begin{quotation}

\noindent \underline{\bf Aphorism 2.} All the anomalies and pathologies of GR and QFT in the form of singularities and other unphysical infinities originate from the {\it a priori} assumption of a background differential manifold as a geometrical model for spacetime.(\bf{F1})

\end{quotation}

Thus, we can combine Aphorisms 1 and 2 to the following `{\em vicious circle}' statement:

\begin{quotation}

\noindent {\bf Fundamental Vicious Circle.} If we wish to formulate the dynamical laws of QG as differential equations proper, it seems that we have to use the concepts and techniques of CDG on a smooth manifold. However, the latter is responsible for both the singularities of GR and the unphysical the infinities of QFT---sites in the spacetime manifold where the laws of physics appear to break down or lead to unphysical infinities for important observable field quantities; hence, we seem to arrive at an impasse.\footnote{The reader should note here that, in the three quotes above, both Feynman and Isham question the {\em Mathematical Efficacy} of CDG in QG (\bf{M1,F1,M2}).}

\end{quotation}

\section{Enter ADG}

Below, in juxtaposition and contradistinction to the points outlined above, we itemise the basic tenets of ADG:

$\bullet$ Abstract Differential Geometry is more of a Leibnizian (:relational), rather than Newtonian (:geometrical), purely algebraic (:sheaf-theoretic and homological algebraic) way of doing differential geometry (:Calculus) \cite{leibniz1,leibniz2}, without at all recourse to or dependence on a pointed, smooth geometric locally Euclidean background spacetime (:a $\smooth$-smooth manifold) for its concepts, technical machinery and constructions thereof \cite{mall1,mall2,mall4}.({\bf M2})

$\bullet$ The most fundamental concept of ADG is that of {\em connection} ({\it viz.} generalised differential) $\conn$ acting on a {\it vector sheaf} $\modl$ over a suitably algebraized, by a certain so-called {\em algebra structure sheaf} $\struc$ of generalised coordinates or what Mallios originally coined `{\em arithmetics} \cite{mall1,mall4}, arbitrary topological space $X$ as a linear and Leibnizian sheaf morphism. The pair $(\conn ,\modl)$ is coined an {\em ADG-field} \cite{mall1,mall4}.({\bf M1,M2})

$\bullet$ Based on the concept of connection, ADG erects the whole edifice of CDG (plus more), but in the manifest absence of a background geometrical manifold \cite{mall1,mall4}. 

$\bullet$ Thus, a new, entirely algebraic (:relational) notion of {\em geometry} emerges, whereby, geometry does not pertain to the configuration states (`shape') and measurement of objects living in an {\it a priori} posited (:postulated) {\em ether-like} background space \cite{df0,df}, but rather, it derives from the algebraic (:dynamical) relations between the objects that live on that `space'.\footnote{These `objects' are the very ADG-connection fields acting on the {\em sections} of the vector sheaves involved.}

$\bullet$ The last bullet point put in more physical terms: {\em physical geometry} (:{\em physical `spacetime'}) is not {\it a priori} posited like the differential spacetime manifold $M$ of GR. Rather, it derives from the dynamics (:the differential equations involving the connection field $\conn$) of the `objects' (:the dynamical physical fields) themselves.\footnote{The reader should refer to the wonderful book \cite{malzaf1} for a thorough technical exposition and semantic explanation of our `motto' here, namely that: {\em geometry and spacetime derive from the ADG-field dynamics, not the other way around}.} ({\bf E2,M2})

We may distill the essential gist of the bullets above to the following Aphorism:

\begin{quotation}

\noindent \underline{\bf Aphorism 3.} From the ADG-theoretic perspective, {\em physical geometry} (or physical `spacetime') derives from, or is the outcome of, the algebraically (:sheaf theoretically) represented dynamical relations between the ADG-fields (:the physical laws, which are formulated categorically as equations between he relevant sheaf morphisms that the ADG-connection fields correspond to).\footnote{In this regard, one may think of the more commonly used mathematical term `{\em solution space}' derived from a set of (differential) equations set up (on a manifold) by the usual CDG-means. That `solution space' {\em is} the `physical geometry' \cite{malzaf1,rap15}.}(\bf{E2,M2})

\end{quotation}

\section{The Point of Pointlessness and Finiteness: the ADG Evasion of Spacetime Singularities and the Management of Infinities}

$\bullet$ In the homological algebraic (:category-theoretic) setting of ADG, the singular, ideal and physically unrealistic notion of a {\em geometrical point}\footnote{The `{\em physically unrealistic}' nature of a geometrical point (:an ideal spacetime event, so to speak) can be appreciated if one considers the fact that one cannot localise an `event' (:measure the value of, say, the gravitational field at a point) with more accuracy than the Planck length without creating a black hole (:think for example of the inner Schwartzschild singularity right at the point-mass source, where the gravitational field blows up without bound.} is meaningless.({\bf F1})

$\bullet$ {\it Mutatis mutandis} for the continuous infinity of point-events that the smooth spacetime manifold accommodates: for example, the non-renormalisable infinities of QFT in Minkowski spacetime effectively arise from the fact that one can in principle pack an uncountable infinity of events (:field values) in a finite spacetime volume.({\bf F1})

$\bullet$ In both the point of pointlessness and finiteness, ADG has been applied towards formulating on the one hand a locally finite, causal and quantal version of Lorentzian vacuum Einstein gravity and free Yang-Mills theories, and on the other, the same dynamical equations are seen to hold over spaces that are everywhere dense with singularities of the most unmanageable and problematic kind from the point of view of CDG \cite{mall3,mall7,mall9,mall11,malrap1,malrap2,malrap3,malros1,malros2,malros3,rap5}.

$\bullet$ {\em The results above are due to the purely algebraic and background pointed continuous spacetime manifold independent character of ADG} \cite{mall1,mall2,mall4,rap14}.

Again, we may distill the essential gist of the bullets above to the following Aphorism, our fourth one:

\begin{quotation}

\noindent \underline{\bf Aphorism 4.} We can formulate the dynamical equations of Einstein and Yang-Mills over highly pathological and problematic spaces, especially when viewed from the smooth background spacetime manifold perspective of CDG. Thus, singularities (and their associated infinities) are not insuperable obstacles and `sites' where the differential equations that represent the dynamical field laws of Nature appear to break down. Not only we can evade them by ADG-theoretic means, but also we can `calculate' (:actually do Calculus!) in their very presence, in spite of them. The inherently algebraic differential geometric mechanism of ADG is genuinely background smooth spacetime independent, hence it does stumble on its inherent anomalies and pathologies  \cite{mall3,mall7,mall9,mall11,malrap1,malrap2,malrap3,malros1,malros2,malros3,rap5}. ({\bf E1,M1,F1,E2,M2})

Thus, by ADG-theoretic means we are able to evade the {\em `vicious circle'} statement that we made earlier after the first two aphorisms, as well as to question both Feynman's (Q2,3) and Isham's (Q4) doubts about using differential geometric ideas in QG research.

\end{quotation}

\subsection{ADG Gravity: Einstein's Purely Algebraic Description of Reality in the Quantum Deep} 

In the Philosophy of Physics there is a well established view that Einstein stubbornly, but in vain, pursued his Unified Field Theory\footnote{Einstein had originally coined it {\em Unitary} Field Theory instead \cite{einst3}.} research on a continuous spacetime manifold in spite of the inherently finitistic and algebraic description of Physical Reality at subatomic scales that Quantum Theory brought about. 

The following couple of quotations from Einstein's The Meaning of Relativity \cite{einst3} testify to that:

\bigskip\noindent (Q5)\hskip 0.9in
\begin{minipage}{11cm}
\noindent ``{\small ...One can give good reasons why reality
cannot at all be represented by a continuous field. {\small\em
From the quantum phenomena it appears to follow with certainty
that a finite system of finite energy can be completely described
by a finite set of numbers}.\footnote{Our emphasis.} This does not
seem to be in accordance with a continuum theory, and must lead to
an attempt to {\small\em find a purely algebraic theory for the
description of reality}\footnote{Our emphasis.}...}''
\end{minipage}

\noindent and a similar quote from \cite{einst3} that also mentions singularities:

\bigskip\noindent (Q6)\hskip 0.9in
\begin{minipage}{11cm}
\noindent ``{\small ...Is it conceivable that a field
theory\footnote{Here, Einstein was implicitly alluding to his
Unitary Field Theory, which, according to his vision, could
hopefully `explain away' quantum phenomena.} permits one to understand the atomistic and quantum
structure of reality? Almost everybody will answer this question
with `no'. But I believe that at the present time nobody knows
anything reliable about it. {\em This is so because we cannot
judge in what manner and how strongly the exclusion of
singularities reduces the manifold of solutions. We do not possess
any method at all to derive systematically solutions that are free
of singularities}\footnote{Our emphasis.}...}'' \cite{einst3}
\end{minipage}

In connection with the two Einstein quotes (Q5,6) above, with our fifth Aphorism next, which closes this section, we kill two birds with one stone as it were:

\begin{quotation}

\noindent \underline{\bf Aphorism 5.} ADG is an entirely algebraic method for formulating gravity and quantum 
Yang-Mills theories of matter field theoretically and finitistically by evading singularities, plus without any dependence on a background geometrical spacetime continuum, with its all inherent singularities and associated unphysical smooth field infinities \cite{malrap3,rap5,rap13,rap14,rap15}.

\end{quotation}

\subsection{Sheaf Theory and the Transition from Local to Global}

At the basis of ADG ({\it alias}, {\em The Geometry of Vector Sheaves}), lie the purely algebraic methods of {\em sheaf theory} \cite{bredon,mall1,mall2,mall4}. Unlike the geometry of smooth vector bundles, which features prominently in the {\em geometrisation of Physics} that gauge theory brought about \cite{gosch}, sheaf theory has been slow in coming in QG research.

Structures closely related to sheaves are special type of categories called {\em topoi} \cite{macmo,maclane2}, which are pointless spaces having their own internal, intuitionistic-type of logic. Topoi have been applied to both quantum logic \cite{buttish3,buttish1,buttish2,buttish4} and quantum spacetime structures, including QG research \cite{ish4,ish3,ish5,ish6,ish7,ish8,rap7,rap11,rap15,malzaf2,malzaf1}.\footnote{We also recommend that the reader refers to Goro Kato's truly original application of sheaf and topos theory to quantum spacetime structure \cite{kato,kato1,kato3,kato4,kato5,kato6}.}

In this regard, very early on, Rudolph Haag \cite{haag} intuited the great import that sheaf theory could bring to QFT as sheaves are structures tailor-cut to encode and transmit information (from local measurements of quantum observables, for example) from local to global in QFT:

\bigskip\noindent (Q7)\hskip 0.9in
\begin{minipage}{11cm}
\noindent ``{\small {\bf Germs.} {\small\em We may take it as the
central message of Quantum Field Theory that all information
characterizing the theory is strictly local i.e. expressed in the
structure of the theory in an arbitrarily small neighborhood of a
point}.\footnote{Our emphasis.} For instance in the traditional
approach the theory is characterized by a Lagrangean density.
{\small\em Since the quantities associated with a point are very
singular objects, it is advisable to consider neighborhoods. This
means that instead of a fiber bundle one has to work with a sheaf.
The needed information consists then of two parts: first the
description of the germs, secondly the rules for joining the germs
to obtain the theory in a finite region}\footnote{Again, emphasis
is ours.}...}''
\end{minipage}

\noindent Indeed, the vector and algebra sheaves involved in ADG and their associated topoi have been used very successfully in analysing the structure of the algebras of local quantum observables and how these stitch up from local to global \cite{zaf1,zaf2}. Moreover, the ADG sheaves and their associated topoi have been applied to address important issues in QG research \cite{rap7,rap11,rap14,malzaf1,malzaf2,rap15}.

\section{Revisiting Feynman: Gravity as Gauge Theory}

Returning to the Feynman quote in Section 2, we wish to dwell a bit on his remark that {\em the fact that the gravitational field was identified with the smooth spacetime metric $g_{\mu\nu}$ of the CDG-based Riemannian geometry on a differential spacetime manifold, was an `accident' of theory making}.\footnote{That is to say, Einstein formulated GR as the dynamics of the metric, `because' he used the CDG-based pseudo-Riemannian geometry of a smooth base spacetime manifold \cite{mtw}.} Rather, Feynman intuited that:

\begin{quotation}
\noindent {\em The deeper character of gravity is that it is a gauge force, much like the other three fundamental forces, while the methods of CDG would be ineffective in the QG deep.}\footnote{Similarly to what Isham said in quote (Q4) following Feynman's (Q2,3).}
\end{quotation}

In subsequent developments in GR, we were able to cast gravity as a gauge theory in the new {\em Ashtekar variables} involving a spin-Lorentzian gravitational connection \cite{ash} and apply the new, first-order formalism\footnote{The Ashtekar formalism in terms of the tetrad $e_{\mu}$ and the spin-connection $\mathcal{A}_{\mu}$ dynamical variables is coined {\em first order}, while the GR of Einstein is based solely on the smooth spacetime metric $g_{\mu\nu}$ as its sole dynamical variable.} to a new candidate for (canonical) QG called {\em Loop Quantum Gravity} \cite{ash0,loop}. Albeit, on the one hand, the metric was still implicitly involved in the dynamics in the guise of the {\em vierbein} comoving tetrad $e_{\mu}$, while the smooth spacetime manifold was still present as a geometrical background in order for the canonical formalism to be applied by the methods of CDG.\footnote{With all the problematic issues and pathologies that this dependence causes to QG (such as {\em the $\mathrm{Diff}(M)$ constraint problem} and {\em the problem of time} \cite{ish2, ish1, ish10}.}

The words in the quote above leave us with the following quandary, posited below as a rhetorical question in the light of ADG:

\begin{quotation}
\noindent {\em Is there a way to view gravity as a gauge theory, as Feynman inuited and envisaged, while still be able to apply to it differential geometric ideas, methods and techniques in spite of Feynman's and Isham's No-Go of CDG in QG research?}
\end{quotation}

\noindent Which brings us to the next subsection about the ADG perspective on gravity as a gauge theory.

\subsection{Enter ADG: Gravity as Pure Gauge Theory of the Third Kind}

One of the central results of the application of ADG to QG is that:

\begin{quotation}
\noindent \underline{\bf Aphorism 6:} {\em Gravity is a pure gauge theory, without recourse to an underlying (smooth) spacetime manifold structure for either its mathematical formulation or its physical interpretation. Gravity involves the dynamics of the ADG-theoretic Einstein field 
$\mathcal{F}_{Einst}=(\modl ,\conn)$, which is simply a connection $\conn$ on a suitable vector sheaf $\modl$. The dynamical Einstein equations are derived from a variational principle applied on the ADG-version of the Einstein-Hilbert action functional} \cite{mall4,malrap3,rap13,rap15}.\footnote{By `{\em pure gauge theory}' above, it is meant that the sole dynamical variable in the theory is the gravitational connection $\conn$, acting on the sections of a suitable vector sheaf $\modl$, and nothing else. The corresponding formalism has been coined {\em half-order formalism}, to distinguish it from the {\em first-order formalism} of Ashtekar and its smooth spin-Lorentzian connection holonomy based Loop QG outgrowth \cite{loop}, and of course from the usual {\em second-order formalism} of the original Einstein theory (GR) \cite{mtw}.} ({\bf E1,M1,F1})
\end{quotation}

\section{Field Solipsism and Functoriality: The Point of Spacetimelessness, Generalised Principle of General Covariance and a Different Perspective on the `Measurement Problem'}

A geometrical point is mathematically an ideal and physically an unrealistic (:singular) entity. We discussed earlier how the sheaf-theoretic ADG and its pointless topos-theoretic extension evade the pointed background geometrical spacetime continuum of events of GR. We also noted how the ADG-gravitational connection field is the sole dynamical variable in the theory, while the underlying spacetime metric $g_{\mu\nu}$ of the usual CDG-based GR is manifestly absent from our theory. This on the one hand is supposed to depict the {\em pure gauge character} of ADG-gravity {\it \`a-la} Feynman, and on the other, to support the aforementioned ADG-field solipsism: that is to say, that 

\begin{quotation}
\noindent  the ADG-gravitational dynamics does not need or depend at all on a background differential spacetime manifold for either its differential geometric formulation as a differential equation proper, or for its physical interpretation. {\em ADG-gravity is a genuinely background independent theory.} The result is that the sole dynamical variable in ADG-gravity is the gravitational Einstein ADG-field $\mathcal{F}_{Einst}=(\modl ,\conn)$, a feature that has been coined {\em field solipsism} and the Einstein-Hilbert variational action principle dynamics that $\mathcal{F}_{Einst}=(\modl ,\conn)$ obeys has been coined {\em ADG-field autodynamics} (:autonomous gravitational field dynamics), with no dependence whatsoever on a background spacetime manifold with its inherent gravitational singularities and unphysical field infinities \cite{malrap3,rap7,rap13,rap14,rap15}.({\bf E1,M1,F1,E2,M2})
\end{quotation}

\noindent The discussion above brings to the forefront one very telling Einstein quote from \cite{einst2}:

\bigskip\noindent (Q8)\hskip 0.9in
\begin{minipage}{11cm}
\noindent ``{\small Time and space are modes by which {\em
we}\footnote{Our emphasis.} think, not conditions in which we
live.}''
\end{minipage}

\noindent Space and time are human inventions convenient for representing, localising and quantifying our measurements of physically observable entities.

Which brings us to the idea of {\em spacetime point coordinates}, or equivalently, {\em spacetime determinations/localisations/measurements of events or field-values}. The locution of every point field-value or event in the spacetime manifold of GR is supposed to be determined (:measured) by four real spacetime coordinates (:coordinate functions) with respect to a given coordinate system (measurement frame of location).

The Principle of General Covariance (PGC) of GR mandates that the law of gravity (:Einstein's equations) is generally covariant; that is to say, it is invariant under any arbitrary general coordinate transformation.\footnote{Technically, we say that the group of symmetries of GR is $GL(4,\R)$, the group of general linear transformations of the locally Euclidean (:$\R^{4}$) spacetime manifold $M$.} Equivalently, we may state the PGC of GR as the Kleinian symmetry group of the background $\smooth$-smooth spacetime manifold $M$, as follows: 

\begin{quotation}
\noindent The symmetry group of GR is the group $\mathrm{Diff}(M)$ of differentiable automorphisms (:diffeomorphisms) of the background smooth spacetime manifold $M$.
\end{quotation}

\noindent By contrast, in ADG-gravity, where we have no smooth background spacetime manifold and {\em the ADG-field autodynamics is purely gauge},

\begin{quotation}
\noindent The gauge symmetry group sheaf is the principal sheaf  $\aut_{\struc}\modl$ of structure sheaf $\struc$-automorphisms of the associated vector sheaf $\modl$ of the ADG-Einstein field $\mathcal{F}_{Einst}=(\modl ,\conn)$ \cite{malrap3,rap7,rap13,rap14,rap15}.
\end{quotation}

\noindent whereby, as noted earlier, $\struc$ is the abelian (:commutative) algebra structure sheaf of generalised arithmetics or coordinates in the theory.

In other words, the ADG-gravitational dynamics, which is formulated entirely categorically in terms of the connection $\conn$ sheaf morphism, is respected by (:`remains invariant under') all our generalised measurements (:arithmetics, event coordinate determinations) encoded in the abelian structure sheaf $\struc$. {\em The gravitational dynamics `sees through' all our coordinate measurements (:spacetime event localisations) in $\struc$} \cite{rap14,rap15}.

\subsection{The Issue of Functoriality}

In \cite{rap15} it has been shown that the aforesaid PGC of GR, which is tantamount to {\em $\struc$-invariance in ADG-gravity}, is an example of {\em the Functoriality of ADG-gravity}. In other words, since the dynamics is categorically represented as equations involving the connection sheaf morphism $\conn$ of the ADG-Einstein field $\mathcal{F}_{Einst}=(\modl ,\conn)$, the PGC is represented by {\em functors} that preserve the relevant categories.\footnote{In category-theoretic parlance, a functor between two categories, is a map or transformation that respects the objects and arrows of the two categories. In \cite{rap15}, in continuation and extension of \cite{mall14,mall15,mall13}, it was shown and argued that the relevant functors are, in fact, special types of $\struc$-preserving functors (or $\struc$-morphisms) called {\em geometric morphisms}, which preserve the `geometric' structure of the vector sheaves involved in the ADG-gravitational field autodynamics.}

We distill all the foregoing discussion into our seventh Aphorism below:

\begin{quotation}

\noindent \underline{\bf Aphorism 7.} In ADG-Gravity, the dynamics is purely gauge and background spacetime manifold independent and functorial, while the PGC is functorially represented in terms of the principal group sheaf of automorphisms of the relevant vector sheaves as $\struc$-invariance \cite{mall14,mall15,mall13,rap15}. The latter simply means that the ADG-field dynamics, which in its Einstein-Hilbert action expression involves the curvature of the connection which is an $\otimes_{\struc}$-tensor, remains invariant (or `unperturbed') by our generalised coordinate `measurements' that are organised in the structure algebra sheaf $\struc$.({\bf E1,M1,F1,E2,M2})

\end{quotation}

\noindent Thus, all our generalised measurements are represented in ADG as sections of the structure algebra sheaf $\struc$.\footnote{The reader should note here that the structure sheaf $\struc$ is supposed to be a sheaf of {\em abelian} (:commutative) algebras. This reflects our primitive assumption that {\em we always measure commutative numbers} (:Bohr's original $c$-numbers), while in the Quantum Theory it is supposed to be the ADG-version of Bohr's Correspondence principle: {\em although quantum observables may be noncommutative $q$-numbers, our measurements thereof are commutative $c$-numbers}.} In turn, the ADG-gravitational autodynamics, since it is $\struc$-functorial,\footnote{Or as Mallios originally coined it: {\em $\struc$-invariant}.} is not `disturbed' at all by our generalised field measurements in $\struc$. Furthermore, the $\otimes_{\struc}$-functorial ADG-gravitational field dynamics does not break down in any differential geometrical sense in the presence of any type of singularity that may be encoded in the structure sheaf $\struc$ of our generalised coordinates (:arithmetics).

The pair of observations above, namely that:

\begin{enumerate}

\item The ADG-gravitational dynamics is unperturbed by our generalised measurements in $\struc$; and,

\item The ADG-gravitational dynamics does not break down in any (differential geometric) sense by any kind of `singularities' or `anomalies' present in $\struc$,

\end{enumerate}

\noindent reflect what we have elsewhere called {\em the Principle of ADG-Field Realism} \cite{rap13,rap14,rap15}. Which brings us to the last section.

\section{Gauge Field Theory of the Third Kind and its Third Quantisation}

The last philosophical issue of ADG-gravity that we would like to discuss in this paper is two-fold: 

\vskip 0.1in

\noindent $\bullet$ {\bf Gauge Field Theory of the Third Kind.} We discussed earlier how from an ADG-theoretic perspective {\em gravity is regarded as a gauge theory}. We noted that the ADG-formalism may be coined {\em half-order formalism}, to distinguish it from the original {\em second-order formalism} of Einstein, whereby the dynamical variable is the smooth Riemannan spacetime metric $g_{\mu\nu}$, or from the more recent {\em first-order formalism} of Ashtekar, whereby the gravitational dynamical variables are the spin-connection $\mathcal{A}_{\mu}$ and the {\em vierbein} comoving frame $e_{\mu}$. In ADG-gravity, the sole dynamical variable is the Einstein (connection) field $\mathcal{F}_{Einst}=(\modl ,\conn)$.

As such, the ADG-based gauge-theoretic formulation of gravity, without recourse to any background spacetime manifold, has been called {\em pure gauge field autodynamics} \cite{malrap3,rap13,rap14,rap15}.

The second denomination, {\em gauge theory of the third kind}, comes from the observation that the first $U(1)$ gauge (or scale) theory for the electromagnetic and the gravitational field due to Weyl \cite{weyl} was a {\em global gauge theory},\footnote{In the sense that Weyl showed that non-spacetime localised (global) gauge/scale invariance implies the conservation of electric charge in much the same way that general coordinate invariance leads to the conservation of energy and momentum in gravitational dynamics. Weyl's original gauge theory is commonly referred to as {\em gauge theory of the 1st kind}.} while the local gauge field theories underlying the three fundamental forces (other than gravity) of the Standard Model\footnote{That is, the electromagnetic (with local gauge group $U(1)$), the weak nuclear (with local gauge group $SU(2)$) and the strong nuclear (with local gauge group $SU(3)$) forces.} are {\em flat Minkowski space localised gauge theories} \cite{gosch}.\footnote{And they are commonly referred to as {\em gauge theories of the 2nd kind}.}

By contrast, aside from its half-order formalism, our ADG-based gauge-theoretic formulation of gravity (:ADG-gravity), although local by its very sheaf-theoretic character, {\em is not background spacetime localised, since there is no background spacetime manifold to localise and solder it on to begin with}.({\bf E1,M1, F1,E2,M2})

\vskip0.1in

\noindent $\bullet$ {\bf Third Quantisation.} In \cite{rap13, rap14}, and recently in \cite{rap15}, a {\em third canonical type of ADG-field quantisation} scenario was proposed according to which certain {\em local, sheaf cohomological characteristic forms} that characterise both the vector sheaf part $\modl$ and the connection part $\conn$ of the ADG-theoretic vacuum Einstein field $\mathcal{F}_{Einst}=(\modl ,\conn)$ were seen to obey canonical (:generalised position-momentum) Heisenberg type commutation relations, albeit, explicitly {\em not} parametrised by a background spacetime manifold.\footnote{In the sense that {\em they are not equal-time commutation relations} which, in the usual canonical QG scenario, would have been required to obey some global hyperbolicity type of foliation of the background spacetime manifold into time-parametrised 3-dimensional spacelike hypersurfaces.} This is to be expected as our ADG-gravity does not depend at all on an external (background) spacetime manifold, as well as to be desired, as our third quantisation scenario would be expected to `algebraically close' within the autonomous and `solipsistic' ADG-theoretic Einstein field $\mathcal{F}_{Einst}=(\modl ,\conn)$. 

We distill these remarks to the following eighth Aphorism:

\begin{quotation}

\noindent \underline{\bf Aphorism 8.} {\em There are no external geometrical structures, such as a background spacetime manifold, in our theory: all there is is $\mathcal{F}_{Einst}=(\modl ,\conn)$ and its Yang-Mills counterparts $\mathcal{F}_{YM}=(\modl ,\conn)$; hence, if the autonomous (:autodynamical) ADG-fields are to be quantum (or quantised) in any way, they should be quantum (or quantised) from within themselves, not from without} \cite{rap13,rap14,rap15}.({\bf E1,M1,F1,E2,M2})

\end{quotation}

\section{Very Brief Philosophical Musings on the Immediate Future of ADG-Gravity}

This author's current research on ADG-gravity focuses on the following three fronts:

\begin{enumerate}

\item To organise the recently discovered `{\em time-asymmetric algebras}' in \cite{rap16}\footnote{These algebras originally appeared, in primitive form, in this author's Ph.D. thesis \cite{rap0}, in which the early seeds for a time-asymmetric quantum spacetime structure and gravity were planted.} into vector sheaves {\it \`a-la} ADG and, by employing the rich differential geometric mechanism of ADG, explore the possibility of developing a time-asymmetric Dirac equation on the resulting sheaves, possibly with ADG-gravity coupled to it \cite{rap18}.({\bf F2})

\item The project above dovetails snugly with our current musings in \cite{rap17}, where we apply ADG to develop a time-asymmetric version of the vacuum Einstein equations for a finitary spin-Lorentzian gravitational connection \cite{rap1,rap2,malrap1,malrap2,malrap3} on Finkelstein's quantum net as originally worked out by Steve Selesnick \cite{sel0,sel1,sel2,sel3,sel4}, and relate this asymmetry to the fundamental asymmetry that Penrose has for many years anticipated that the true QG theory should account for \cite{jacobs}.({\bf F2})

\item The main philosophical query that will arise from the three papers above \cite{rap16,rap17,rap18} is that the anticipated fundamental time-asymmetry of the true QG theory may not only be traced back to time-asymmetric initial conditions for the Universe,\footnote{Like Penrose's {\em Weyl curvature hypothesis}.} but also it may be due to the fundamentally time-asymmetric quantum gravitational dynamics themselves (:time-asymmetric vacuum Einstein equations for ADG-gravity).({\bf F2})

\end{enumerate}

\noindent {\bf The quest continues.}

\section*{Acknowledgments}

I am greatly indebted to {\em Professors Goro Kato} (Department of Mathematics, California Polytechnic Institute, San Luis Obispo) and {\em Steve Selesnick} (Department of Mathematics, University of St Louis, Missouri) for numerous stimulating exchanges on a plethora of topics in Mathematics, Physics, Philosophy and Poetry after a long hiatus period of personal reflection and research course re-evaluation and re-adjustment.

\hskip0.02in

\centerline{$<.><.><.><.><.><.><.><.><.>$}

\hskip0.02in

The present paper is lovingly dedicated to my parents, {\em George} and {\em Helen Raptis}, whose unceasing moral and material support of my research quests has never been diminished by the passage of time, no matter what its toll on both their ageing bodies and their lucid minds.

\hskip0.02in

\centerline{$<.><.><.><.><.><.><.><.><.>$}

\hskip0.02in

\noindent Last but not least, the unceasing `moral' support of my lovely family: {\em Kathleen, Francis, James} and {\em Cookie}, is also warmly aknowledged, especially their patience and understanding in putting up with me over the years.


\begin{thebibliography}{99}

\bibitem{arist} Aristotle, {\itshape The Nicomachean Ethics}, Oxford World's Classics, Oxford University Press (2009).

\bibitem{ash} Ashtekar, A., {\itshape New Variables for Classical
and Quantum Gravity}, Physical Review Letters, {\bf 57}, 2244
(1986).

\bibitem{ash0} Ashtekar, A., {\itshape Quantum Gravity: What and
Why?}, in {\itshape Asymptotic Quantization}, a volume based on
the author's 1984 Naples Lectures, Monographs and Textbooks in
Physical Science, Lecture Notes {\bf 2}, Bibliopolis, Naples
(1987).

\bibitem{ash1} Ashtekar, A., {\itshape Quantum Gravity: A
Mathematical Physics Perspective}, pre-print (1994);
hep-th/9404019.

\bibitem{au} Auyang, S. Y., {\itshape How is Quantum Field Theory
Possible?}, Oxford University Press, New York-Oxford (1995).

\bibitem{bomb87} Bombelli, L., Lee, J., Meyer, D. and Sorkin, R.
D., {\itshape Space-Time as a Causal Set}, Physical Review
Letters, {\bf 59}, 521 (1987).

\bibitem{bredon} Bredon, G. E., {\itshape Sheaf Theory}, McGraw-Hill, New York (1967).

\bibitem{leibniz2} Brown, R. C., {\itshape The Tangled Origins of the Leibnizian Calculus: A Case Study of a Mathematical Revolution}, World Scientific (2012).

\bibitem{buttish3} Butterfield, J., Hamilton, J. and Isham, C. J., {\itshape A
Topos Perspective on the Kochen-Specker Theorem: III. Von Neumann
Algebras as the Base Category}, International Journal of
Theoretical Physics, {\bf 39}, 2667 (2000).

\bibitem{buttish1} Butterfield, J. and Isham, C. J., {\itshape A Topos
Perspective on the Kochen-Specker Theorem: I. Quantum States as
Generalized Valuations}, International Journal of Theoretical
Physics, {\bf 37}, 2669 (1998).

\bibitem{buttish2} Butterfield, J. and Isham, C. J., {\itshape A Topos
Perspective on the Kochen-Specker Theorem: II. Conceptual Aspects
and Classical Analogues}, International Journal of Theoretical
Physics, {\bf 38}, 827 (1999).

\bibitem{buttish4} Butterfield, J. and Isham, C. J.,
{\itshape Some Possible Roles for Topos Theory in Quantum Theory
and Quantum Gravity}, Foundations of Physics, {\bf 30}, 1707
(2000).

\bibitem{dirac3} Dirac, P. A. M., {\itshape Quantized Singularities in the Electromagnetic Field}, Proceedings of the
Royal Society London {\it A}, {\bf 133}, 60 (1931).

\bibitem{leibniz1} Edwards, C. H. Jr, {\itshape The Historical Development of the Calculus}, 
Springer-Verlag, Berlin-Heidelberg-New York (1982).

\bibitem{einst2} Einstein, A., {\itshape Albert Einstein:
Philosopher-Scientist}, The Library of Living Philosophers, {\bf
7}, Schilpp, P. A. (Ed.), Evanston, III (1949).

\bibitem{einst3} Einstein, A., {\itshape The Meaning of Relativity}, 5th edition,
Princeton University Press, Princeton (1956).

\bibitem{feyn3} Feynman, R. P., {\itshape The Character of Physical Law}, Penguin
Books, London (1992).

\bibitem{feyn} Feynman, R. P., {\itshape Feynman Lectures on
Gravitation}, notes by Morinigo, F. B. and Wagner, W. G.,
Hatfield, B. (Ed.), Penguin Books, London (1999).

\bibitem{df0} Finkelstein, D., {\itshape Theory of Vacuum} in
{\itshape The Philosophy of Vacuum}, Saunders, S. and Brown, H.
(Eds.), Clarendon Press, Oxford (1991).

\bibitem{df} Finkelstein, D. R., {\itshape Quantum Relativity: A Synthesis of the Ideas of Einstein
and Heisenberg}, Springer-Verlag, Berlin-Heidelberg-New York
(1996).

\bibitem{geroch1} Geroch, R., {\itshape What is a singularity in
General Relativity?}, Annals of Physics, {\bf48}, 526 (1968).

\bibitem{geroch2} Geroch, R., {\itshape Local Characterization of Singularities in General
Relativity}, Journal of Mathematical Physics, {\bf 9}, 450 (1968).

\bibitem{gosch} G\"{o}ckeler, M. and Sch\"{u}cker, T., {\itshape
Differential Geometry, Gauge Theories and Gravity}, Cambridge
University Press, Cambridge (1990).

\bibitem{gsw} Green, M. B., Schwartz J. H. and Witten, E., {\itshape Superstring Theory: Introduction (Volume 1)}, Cambridge University Press (2012).\footnote{Online publication of the original book.}

\bibitem{haag} Haag, R., {\itshape Local Quantum Physics: fields, particles,
algebras}, 2nd edition, Springer-Verlag, Berlin-Heidelberg-New
York (1996).

\bibitem{hawk1} Hawking, S. W., {\itshape Singularities in the universe},
Physical Review Letters, {\bf 17}, 444 (1966).

\bibitem{hawk2} Hawking, S. W., {\itshape Breakdown of predictability in gravitational collapse},
Physical Review, {\bf D14}, 2460 (1976).

\bibitem{hawk3} Hawking, S. W. and Ellis, G. F. R., {\itshape The
Large Scale Structure of Space-Time}, Cambridge University Press,
Cambridge (1973).

\bibitem{hawk4} Hawking, S. W. and Penrose, R., {\itshape The Singularities of
Gravitational Collapse and Cosmology}, Proceedings of the Royal
Society London A, {\bf 314}, 529 (1970).

\bibitem{ish} Isham, C. J., {\itshape Canonical groups and the quantization of
geometry and topology}, in {\itshape Conceptual Problems of
Quantum Gravity}, Ashtekar, A. and Stachel, J. (Eds.),
Birkh\"{a}user, Basel (1991).

\bibitem{ish2} Isham, C. J., {\itshape Canonical Quantum Gravity and the Problem of Time}, in {\itshape Integrable Systems,
Quantum Groups, and Quantum Field Theories}, Kluwer Academic
Publishers, London-Amsterdam (1993); gr-qc/9210011.

\bibitem{ish1} Isham, C. J., {\itshape Prima Facie Questions in
Quantum Gravity}, pre-print (1993); gr-qc/9310031.

\bibitem{ish10} Isham, C. J., {\itshape Structural Issues in
Quantum Gravity}, Plenary Session Lecture given at the GR14
Conference (Florence, Italy), pre-print (1995); gr-qc/9510063.

\bibitem{ish4} Isham, C. J., {\itshape Topos Theory and Consistent
Histories: The Internal Logic of the Set of All Consistent Sets},
International Journal of Theoretical Physics, {\bf 36}, 785
(1997).

\bibitem{ish3} Isham, C. J., {\itshape Some Reflections on the Status of Conventional Quantum Theory when Applied to
Quantum Gravity}, in {\itshape The Future of Theoretical Physics
and Cosmology: Celebrating Stephen Hawking's 60th Birthday},
Gibbons, G. W., Shellard, E. P. S. and Rankin, S. J. (Eds.),
Cambridge University Press, Cambridge (2003); quant-ph/0206090.

\bibitem{ish5} Isham, C. J., {\itshape A new approach to quantising space-time:
I. Quantising on a general category}, Advances in Theoretical and
Mathematical Physics, {\bf 7}, 331 (2003); gr-qc/0303060.

\bibitem{ish6} Isham, C. J., {\itshape A new approach to quantising space-time:
II. Quantising on a category of sets}, Advances in Theoretical and
Mathematical Physics, {\bf 7}, 807 (2004); gr-qc/0304077.

\bibitem{ish7} Isham, C. J., {\itshape A new approach to quantising space-time:
III. State vectors as functions on arrows}, Imperial College
pre-print TP/2-03/15 (2003); gr-qc/0306064.

\bibitem{ish8} Isham, C. J., {\itshape Quantising on a category}, to appear in
{\itshape A Festschrift for James Cushing} (2005);
quant-ph/0401175.

\bibitem{jacobs} Jacobs, T. and Maes, C., {\itshape Reversibility and Irreversibility within the Quantum Formalism}, pre-print (2005); quant-ph/0508041.

\bibitem{kato} Kato, G., {\itshape Elemental Principles of Temporal Topos}, Europhysics
Letters, {\bf 68}, 467 (2004).

\bibitem{kato1} Kato, G., {\itshape Presheafification of Time, Space and Matter}, pre-print (2004) (in preparation,
to be submitted to Europhysics Letters).\footnote{Pre-print
available.}

 \bibitem{kato3} Kato, G., {\itshape Elemental Principles of Relativistic t-Topos}, Europhysics Letters, {\bf 71},  (2005).

\bibitem{kato4} Kato, G., {\itshape Kato, G., Microcosm to Macrocosm via the Notion of a Sheaf (Observers in terms of t-Topos)}, in Physics and the Emergence of Organisation, pp. 229-232, World Scientific (2008).

\bibitem{kato5} Kato, G., {\itshape u-Singularity and t-Topos Theoretic Entropy}, International Journal of Theoretical Physics, {\bf 49}, 1952 (2010). 

\bibitem{kato6} Kato, G., {\itshape Elements of Temporal Topos}, Arima Publishing Co. (2013).

\bibitem{kiefer} Kiefer, K., {\itshape On a Quantum Weyl Curvature Hypothesis}, pre-print (2022); gr-qc/2111.02137v2.

\bibitem{leibniz} Leibniz, G. W., {\itshape Discourse on Metaphysics and the Monadology}, translated
by Montgomery, G. R., Great Books in Philosophy Series, Prometheus
Books, Amherst, New York (1992).

\bibitem{loop} Lu, L. and May, P. A., {\itshape Step-by-Step Canonical Quantum Gravity:
Part I: Ashtekar’s New Variables}, pre-print (2024); gr-qc/2401.06863v1.

\bibitem{maclane2} MacLane, S., {\itshape Categories for the Working Mathematician}, Graduate Texts in
Mathematics Series, Springer-Verlag, New York (1971).

\bibitem{macmo} MacLane, S. and Moerdijk, I., {\itshape
Sheaves in Geometry and Logic: A First Introduction to Topos
Theory}, Springer-Verlag, New York (1992).

\bibitem{mall1} Mallios, A.,
{\itshape Geometry of Vector Sheaves: An Axiomatic Approach to
Differential Geometry}, vols. 1-2, Kluwer Academic Publishers,
Dordrecht (1998).\footnote{There is also a Russian translation of
this 2-volume book by MIR Publishers, Moscow (vol. 1, 2000 and
vol. 2, 2001).}

\bibitem{mall2} Mallios, A., {\itshape On an Axiomatic Treatment
of Differential Geometry via Vector Sheaves. Applications}, Math.
Jap. (International Plaza), {\bf 48}, 93 (1998). (invited paper)

\bibitem{mall5} Mallios, A., {\itshape On an axiomatic approach to
geometric prequantization: A classification scheme {\it \`a la}
Kostant-Souriau-Kirillov}, J. Math. Sci. (NY), {\bf 95}, 2648
(1999). (invited paper)

\bibitem{mall3} Mallios, A., {\itshape Abstract Differential Geometry,
General Relativity and Singularities}, in {\itshape Unsolved
Problems in Mathematics for the 21st Century: A Tribute to Kiyoshi
Is\'{e}ki's 80th Birthday}, Abe, J. M. and Tanaka, S. (Eds.), 77,
IOS Press, Amsterdam (2001). (invited paper)

\bibitem{mall7} Mallios, A., {\itshape Remarks on
``singularities''} (2002) (pre-print); gr-qc/0202028.

\bibitem{mall0} Mallios, A., {\itshape On Localizing Topological Algebras}, in {\it Topological Algebras and Their Applications}, Arizmendi, H., Bosch, C., and Palacios, L. (Eds), Contemporary Mathematics, AMS, {\bf 341} (2004) (pre-print); gr-qc/0211032.

\bibitem{mall4} Mallios, A., {\itshape Modern Differential Geometry in Gauge Theories. Vol.I: Maxwell
Fields, Vol.II: Yang-Mills Fields}, 2-volume continuation
(including abstract integration theory) of \cite{mall1},
Birkh\"auser, Boston-Basel-Berlin (Vol. I 2005, Vol. II 2006).

\bibitem{mall6} Mallios, A., {\itshape K-Theory of topological algebras
and second quantization}, Acta Universitatis Ouluensis--Scientiae
Rerum Naturalium, {\bf A408}, 145 (2004); math-ph/0207035.

\bibitem{mall9} Mallios, A., {\itshape Abstract Differential Geometry, Singularities and Physical
Applications}, in {\itshape Topological Algebras with Applications
to Differential Geometry and Mathematical Physics}, in {\sl
Proceedings of a Fest-Colloquium in Honour of Professor Anastasios
Mallios (16--18/9/1999)}, Strantzalos, P. and Fragoulopoulou, M.
(Eds.), Department of Mathematics, University of Athens
Publications (2002).

\bibitem{mall11} Mallios, A., {\itshape Quantum gravity and ``singularities''},
Note Mat., {\bf 25}, 57 (2006) (invited paper); physics/0405111.

\bibitem{mall10} Mallios, A., {\itshape Geometry and physics today}, in a Special Proceedings issue for {\itshape
Glafka--2004: Iconoclastic Approaches to Quantum Gravity}, Raptis,
I. (Ed.),  Int. J. Theor. Phys. {\bf 45}, 1552 (2006) (invited paper); physics/0405112.

\bibitem{mall14} Mallios, A., {\itshape $\mathcal{A}$-Invariance: An axiomatic approach to quantum relativity}, Int. J. Theor. Phys., {\bf 47}, 1929 (2008).

\bibitem{mall15} Mallios, A., {\itshape Relational mathematics: A response to quantum gravity}, Publications Ecole Norm. Sup\'{e}r., Takaddoum, Rabat, Morocco 2007, pp. 61-68 (2010).

\bibitem{mall13} Mallios, A., {\itshape On Utiyama's Theme Through ``$\mathcal{A}$-Invariance''}, Complex Analysis and Operator Theory, {\bf 6}, 775 (2012).

\bibitem{mall12}  Mallios, A., {\itshape Bohr's correspondence principle (:the commutative substance of the quantum), abstract (:axiomatic) quantum gravity, and functor categories}, Manuscript/pre-print (2012).

\bibitem{malrap1} Mallios, A. and Raptis, I.,
{\itshape Finitary Spacetime Sheaves of Quantum Causal Sets:
Curving Quantum Causality}, Int. J. Theor. Phys., {\bf 40}, 1885
(2001); gr-qc/0102097.

\bibitem{malrap2} Mallios, A. and Raptis, I., {\itshape Finitary \v{C}ech-de Rham
Cohomology}, Int. J. Theor. Phys., {\bf 41}, 1857 (2002);
gr-qc/0110033.

\bibitem{malrap3} Mallios, A. and Raptis, I., {\itshape Finitary, Causal and Quantal Vacuum Einstein Gravity},
Int. J. Theor. Phys., {\bf 42}, 1479 (2003); gr-qc/0209048.

\bibitem{malrap4} Mallios, A. and Raptis, I., {\itshape $\smooth$-Smooth
Singularities Exposed: Chimeras of the Differential Spacetime
Manifold}, research monograph (2005) (in preparation);
gr-qc/0411121.\footnote{Two years' old version posted at gr-qc.}

\bibitem{malros1} Mallios, A. and Rosinger, E. E., {\itshape
Abstract Differential Geometry, Differential Algebras of
Generalized Functions and de Rham Cohomology}, Acta Appl. Math.,
{\bf 55}, 231 (1999).

\bibitem{malros2} Mallios, A. and Rosinger, E. E., {\itshape
Space-Time Foam Dense Singularities and de Rham Cohomology}, Acta
Appl. Math., {\bf 67}, 59 (2001).

\bibitem{malros3} Mallios, A. and Rosinger, E. E., {\itshape Dense Singularities and de Rham Cohomology},
in {\itshape Topological Algebras with Applications to
Differential Geometry and Mathematical Physics}, in {\sl
Proceedings of a Fest-Colloquium in Honour of Professor Anastasios
Mallios (16--18/9/1999)}, Strantzalos, P. and Fragoulopoulou, M.
(Eds.), Department of Mathematics, University of Athens
Publications (2002).

\bibitem{malzaf2} Mallios, A. and Zafiris, E., {\itshape Topos-theoretic Relativization of Physical Representability and Quantum Gravity}, pre-print (2007); gr-qc/0610113.

\bibitem{malzaf1} Mallios, A. and Zafiris, E., {\itshape Differential Sheaves and Connections: 
A Natural Approach to Physical Geometry}, Series on Concrete and Applicable Mathematics, {\bf Vol. 18}, World Scientific (2015).

\bibitem{mtw} Misner, C. W., Thorne, K. and Wheeler, J. A.,
{\itshape Gravitation}, Freeman Publishers, San Francisco (1973).

\bibitem{rap0} Raptis, I.,
{\itshape Axiomatic Quantum Timespace Structure: A Preamble to the Quantum Topos Conception of the (Minkowski) Vacuum}, Theoretical Physics Group, Physics Department, The University of Newcastle upon Tyne, United Kingdom (1998).

\bibitem{rap1} Raptis, I.,
{\itshape Algebraic Quantization of Causal Sets}, Int. J. Theor.
Phys., {\bf 39}, 1233 (2000); gr-qc/9906103.

\bibitem{rap2} Raptis, I.,
{\itshape Finitary Spacetime Sheaves}, Int. J. Theor. Phys., {\bf
39}, 1703 (2000); gr-qc/0102108.

\bibitem{rap5}  Raptis, I., {\itshape Finitary-Algebraic `Resolution' of the
Inner Schwarzschild Singularity}, Int. J. Theor. Phys., {\bf 45},
(6) (2006) (to appear); gr-qc/0408045.

\bibitem{rap7} Raptis, I., {\itshape Finitary Topos for Locally Finite, Causal
and Quantal Vacuum Einstein Gravity},  Int. J. Theor. Phys., {\bf 46}, 688 (2007); gr-qc/0507100.

\bibitem{rap11} Raptis, I., {\itshape `Iconoclastic' Categorical Quantum Gravity}, published in a Special Proceedings issue for
{\itshape Glafka--2004: Iconoclastic Approaches to Quantum
Gravity}, Raptis, I. (Ed.), Int. J. Theor. Phys., {\bf 45}, 1495 (2006);
gr-qc/0509089.

\bibitem{rap13} Raptis, I., {\itshape `Third' Quantization of Vacuum Einstein Gravity and Free Yang-Mills Theories}, Int. J. Theor. Phys., {\bf 46}, 1137 (2007); gr-qc/0606021.

\bibitem{rap14} Raptis, I., {\itshape A Dodecalogue of Basic Didactics from Applications of Abstract Differential Geometry to Quantum Gravity}, Int. J. Theor. Phys., {\bf 46}, 3009 (2007); gr-qc/0607038.

\bibitem{rap15} Raptis, I., {\itshape Functoriality in Finitary Vacuum Einstein Gravity and Free Yang-Mills Theories from an Abstract Differential Geometric Perspective}, invited paper contribution to a Special Memorial Volume titled {\em Physical Geometry: Unravelling the Weave of Quantum Geometry} in memory of Professor Anastasios Mallios, edited by Dr Elias Zafiris (Editor-in-Chief), Technical University of Vienna Press (September 2024); gr-qc/2401.13283v3 (vol. 3).

\bibitem{rap16} Raptis, I., {\itshape Multiplicatively Ordered and Directed Hybrid $\delta$-Jordan-Lie Superalgebra}, paper submitted to the Journal of Algebra and its Applications (2024); math-ph/2405.01181v3 (vol. 3).

\bibitem{rap17} Raptis, I., {\itshape ``{\bf Gravity from Light}''. Gravity and Electrodynamics Viewed in the Purely Gauge-Theoretic Light of Abstract Differential Geometry: Tracing the sub-Planckian Origins of Inertial Mass}, pre-print in preparation (August-October 2024); gr-qc/??? (to be posted duly at the e-archives).

\bibitem{rap18} Raptis, I., {\itshape Three Potential Physical Applications of a Hybrid $\delta$-Jordan-Lie Superalgebra: Extended Supersymmetry, Time-Irreversible Free Generative Linear Semigroups, and the Time-Asymmetric Dirac Equation}, pre-print in preparation (August-October 2024); gr-qc/???  (to be posted duly at the e-archives).

\bibitem{rapzap1} Raptis, I. and Zapatrin, R. R.,
{\itshape Quantization of discretized spacetimes and the
correspondence principle}, Int. J. Theor. Phys., {\bf 39}, 1
(2000); gr-qc/9904079.

\bibitem{rapzap2} Raptis, I. and Zapatrin, R. R.,
{\itshape Algebraic description of spacetime foam}, Class. Quant.
Grav., {\bf 20}, 4187 (2001); gr-qc/0102048.

\bibitem{sel0} Selesnick, S. A, {\itshape Second quantization, projective modules and local gauge invariance},  Int. J. Theor. Phys., {\bf 22}, 29 (1983).

\bibitem{sel1} Selesnick, S. A., {\itshape Gauge fields on the quantum net}, Journal of Mathematical Physics, {\bf 36}, 5465 (1995).

\bibitem{sel2} Selesnick, S. A., {\itshape Quanta, Logic and Spacetime}, World Scientific (2003).

\bibitem{sel3} Selesnick, S. A., {\itshape Computing the Lagrangians of the Standard Model II. The Ghost Term}, Int. J. Theor. Phys., {\bf 55}, 4999 (2016).

\bibitem{sel4} Selesnick, S. A., {\itshape Emergent geometry, emergent forces}, Class. Quant.
Grav., {\bf 34}, 195010 (2017).

\bibitem{senzwiebach} Sen A. and Zwiebach B., {\itshape String Field Theory: A Review}, pre-print (2024); gr-qc/2405.19421v2.

\bibitem{sork4} Sorkin, R. D., {\itshape Does a Discrete Order Underlie Spacetime and its Metric?} in
{\itshape Proceedings of the Third Canadian Conference on General
Relativity and Relativistic Astrophysics}, Cooperstock, F. and
Tupper, B. (Eds.), World Scientific, Singapore (1990).

\bibitem{sork1} Sorkin, R. D.,
{\itshape A Specimen of Theory Construction from Quantum Gravity},
in {\itshape The Creation of Ideas in Physics}, Leplin, J. (Ed.),
Kluwer Academic Publishers, Dordrecht (1995); gr-qc/9511063.

\bibitem{sork2} Sorkin, R. D., {\itshape Forks in the Road, on the Way to
Quantum Gravity}, International Journal of Theoretical Physics,
{\bf 36}, 2759 (1997); gr-qc/9706002.

\bibitem{stachel} Stachel, J. J., {\itshape The Other Einstein:
Einstein Contra Field Theory}, in {\itshape Einstein in Context},
Beller, M., Cohen, R. S. and Renn, J. (Eds.), Cambridge University
Press, Cambridge (1993).

\bibitem{thiem2} Thiemann, T., {\itshape Introduction to Modern Canonical Quantum General Relativity},
pre-print (2001); gr-qc/0110034.

\bibitem{thooft} 't Hooft, G., {\itshape Obstacles on the Way
Towards the Quantization of Space, Time and Matter},
ITP-University of Utrecht, pre-print SPIN-2000/20 (2001).

\bibitem{vas1} Vassiliou, E., {\itshape On Mallios'
$\aconn$-connections as connections on principal sheaves}, Note di
Matematica, {\bf 14}, 237 (1994).

\bibitem{vas2} Vassiliou, E., {\itshape Connections on principal
sheaves}, in {\itshape New Developments in Differential Geometry},
Szenthe, J. (Ed.), Kluwer Academic Publishers, Dordrecht (1999).

\bibitem{vas3} Vassiliou, E., {\itshape On the geometry of
associated sheaves}, Bulletin of the Greek Mathematical Society,
{\bf 44}, 157 (2000).

\bibitem{vas4} Vassiliou, E., {\itshape Geometry of Principal Sheaves}, Mathematics and Its Applications, {\bf Vol. 578}, Springer (2005).

\bibitem{weyl} Weyl, H., {\itshape Gravitation and Electricity}, in {\itshape The Principle of Relativity},
Dover Publications, New York (1952).

\bibitem{wheeler1} Wheeler, J. A., {\itshape Singularity and
Unanimity}, General Relativity and Gravitation, {\bf 8}, 713
(1977).

\bibitem{zaf1} Zafiris, E., {\itshape Quantum observables algebras and abstract differential geometry}, Int. J. Theor. Phys., {bf 46}, 319 (2007).

\bibitem{zaf2} Zafiris, E., {\itshape Physical Principles of Functorial Gauge Localization and Dynamics. With a View Toward Quantum General Relativity.}, Monograph in preparation (2012).

\end{thebibliography}
\end{document}